\documentclass{article}

\usepackage{amsfonts}
\usepackage{amsmath}
\usepackage{amssymb}
\usepackage{amsthm}

\newtheorem{thm}{Theorem}[section]

\newtheorem{lem}[thm]{Lemma}

\theoremstyle{definition}

\newtheorem{rem}[thm]{Remark}

\begin{document}

\title{An active attack on a distributed Group Key Exchange system\thanks{ Second and fourth author are partially
suppported by  Ministerio de Economia y
Competitividad grant MTM2014-54439 and Junta de Andalucia (FQM0211).
Third author is partially supported by \emph{Armasuisse} and Swiss National Science
Foundation grant number 149716.
} } 

\author{M. Baouch\footnote{University of Almeria}, J.A.~L\'opez-Ramos\footnotemark[2],  R.~Schnyder\footnote{University of Zurich}, B. Torrecillas\footnotemark[2]
}

\maketitle

\begin{abstract}
In this work, we introduce an active attack on a Group Key Exchange
protocol by Burmester and Desmedt. The attacker obtains a copy of the
shared key, which is created in a collaborative manner with the legal
users in a communication group.
\end{abstract}

\section{Introduction}

Group Key Exchange (GKE) has recently been a concern mainly due to
the huge development of multiparty communications that, nowadays,
are applied in many networks and, in most cases, with a very light
infrastructure. For this reason, distributed GKE, where members in a
group collaborate to agree on a common key, is becoming very popular
and there exist many approaches trying to provide effective
protocols to this end (cf. \cite{lee} or \cite{vandenmerwe} for
example).

Some efficient solutions were introduced by Burmester and Desmedt in
\cite{burmester1} and \cite{burmester2} and by Steiner et al. in
\cite{steiner1} and \cite{steiner2} that extend naturally the
classical Diffie-Hellman protocol (\cite{diffie}). Both solutions
were shown to be secure against a passive adversary if the
Diffie-Hellman problem is intractable. However, in \cite{schnyder}
the authors provide an active attack on one of Steiner et al.'s
proposals that allows an intrusion into the communicating group,
assuming control of communications of two particular parties only
during the key exchange.

Motivated by this work, which exploits a weakness of the protocol
consisting of the possibility to ask one of the users like an
oracle, we show a similar active attack on Burmester and Desmedt's
proposal (\cite{burmester1} and \cite{burmester2}) that presents a
similar weakness. In this case, our attack requires control of the
communications of only one user and, as in the case of
\cite{schnyder}, for only the duration of the key exchange, which is
to say that after the attack, the attacker does not need to control
communications of this user to translate messages, since all users
and the attacker him/herself agree on a common key. We also note
that since rekeying operation in this case is carried out by
rerunning the protocol completely, the attacker can repeat the
strategy (not necessarily on the same user) and keep listening to
all communications for an unlimited time.

The following sections describe the protocol introduced in
\cite{burmester1} and \cite{burmester2} in an algebraic group
setting and the active attack respectively.

\section{The Group Key Exchange protocol}

Let $U_i$, $i=1, \dots, n$ be a set of parties that want to generate
a shared key $K$. Let $G$ be a group of prime order $q$. The users
agree on a generator $g$ of $G$ and operate as follows:

\medskip

\noindent {\bf Round 1.} Each party $U_i$, $i=1, \dots ,n$, selects
a random $r_i \in \mathbb{Z}_q$ and broadcasts $z_i=g^{r_i}$.

\medskip

\noindent {\bf Round 2.} Each party $U_i$, $i=1, \dots ,n$,
broadcasts $X_i=(z_{i+1}/z_{i-1})^{r_i}$.

\medskip

\noindent {\bf Key Computations.} Each party $U_i$, $i=1, \dots ,n$
computes the key
\[K_i=(z_{i-1})^{nr_i} \cdot X_i^{n-1} \cdot X_{i+1}^{n-2} \cdots
X_{n+i-2} \in G.\]

\medskip

In the above, indices should be interpreted modulo $n$.
By \cite[Lemma 3.1]{burmester2}, the users $U_i$, $i=1, \dots ,n$
compute the same key $K=g^{r_1r_2+r_2r_3+\cdots +r_nr_1} \in G$.

\section{The Attack}

Under the conditions of the previous sections, let $U_i$, $i=1,
\dots, n$, be a set of communicating parties and let $A$ be an active
attacker that is able to take control of one of the users'
communications, let us say $U_k$. Then the attack is developed as
follows.

\begin{enumerate}

\item Each party $U_i$, $i=1, \dots ,n$, selects a random $r_i \in
\mathbb{Z}_q$ and broadcasts $z_i=g^{r_i}$ as in round 1 of the
protocol.

\item $A$ stops $r_k$ and, forging $U_k$'s identity, sends to $U_i$,
$i=1, \dots ,n$, $i\not= k$, $z'_k=g^a$, where $a$ is such that
$a-1$ is invertible in $\mathbb{Z}_q$.

\item At the same time, $A$ stops the message $z_{k+1}$ for $U_k$
and replaces it by $z'_{k+1}=z_{k-1}^a=(g^{r_{k-1}})^a$.

\item $U_k$ starts round 2 and computes
$X_k=(z'_{k+1}/z_{k-1})^{r_k}=(z_{k-1}^{r_k})^{a-1}$, which is
broadcasted.

\item $A$ stops $X_k$ and $U_k$ is waiting in round 2 to receive the
remaining $X_i$, $i=1, \dots ,n$, $i\not= k$.

\item While $X_k$ is waiting in round 2, $A$ finishes running the
GKE protocol with participants $U_i$, $i=1, \dots ,n$, $i\not= k$,
using $A$'s private information $a$. They agree on a key $K$.

\item $A$ computes $b=(a-1)^{-1}\ \mbox{mod} \ q$ and computes
$X_k^b=z_{k-1}^{r_k}$.

\item $A$ generates a list $\{ h_1, \dots , h_{n-3} \}$ of elements
in $G$ and provides $U_k$ the list $\{ X_1, \dots , X_{k-1},
X_{k+1}, \dots , X_n \}$ given by
\begin{align*}
X_{k+1} &=  z_{k-1}^{-r_k}\ h_1, \\
X_{k+j} & =  h_{j-1}^{-1}h_j, \ \mbox{for} \ j=2, \dots, n-3, \\
X_{k-2} & =  K \ X_k^{-(n-1)} \ z_{k-1}^{-2r_k}\ h_{n-3}^{-2}\
\textstyle\prod_{r=1}^{n-4}h_r^{-1},
\end{align*}
where indices are again taken modulo $n$.

\end{enumerate}

\begin{rem}
Let us note that $X_{k-1}$ could be any arbitrary element since this
is not used to compute $K_{k-1}$. However, in a proper execution of
the protocol, it holds that $\prod_{i=1}^nX_i=1$. User $U_k$ could
check whether this holds. In order to avoid being detected, once we
have computed all $X_i$ with $i\not= 1$, we can define $X_{k-1} =
(\prod_{i=1, i\not=k-1}^nX_i)^{-1}$.
\end{rem}

\begin{lem}
After the active attack, all users $U_i$, $i=1, \dots ,n$, and $A$
share the same key.
\end{lem}

\noindent{\it Proof.} It is clear from step 6 that $A$ and $U_i$,
$i=1, \dots ,n$, $i\not= k$ share the key~$K$. A straightforward
computation shows that \[K_k=(z_{k-1})^{nr_k} \cdot X_k^{n-1} \cdot
X_{k+1}^{n-2} \cdots X_{n+k-2}=K.\]

\end{document}